\documentstyle[12pt,aasms4]{article}
\def\lapprox{\hbox{\lower .8ex\hbox{$\,\buildrel < \over\sim\,$}}}
\def\gapprox{\hbox{\lower .8ex\hbox{$\,\buildrel > \over\sim\,$}}}

\begin{document}

\title
{Further constraints on white dwarf galactic halos}

\author
{R. Canal\altaffilmark{1}, J. Isern \altaffilmark{2}, and P.
Ruiz--Lapuente\altaffilmark{1,3}}

\altaffiltext
{1}{Department of Astronomy, University of Barcelona, Mart\'\i\ i Franqu\'es
1, E--08028 Barcelona, Spain. E--mail: ramon@mizar.am.ub.es,
pilar@mizar.am.ub.es}

\altaffiltext
{2}{Institut d'Estudis Espacials de Catalunya, Edifici Nexus--104, Gran
Capit\`a 2--4, E--08034 Barcelona, Spain. E--mail: isern@ieec.fcr.es}

\altaffiltext
{3}{Max--Planck--Institut f\"ur Astrophysik, Karl--Schwarzschild--Strasse 1,
D--85740 Garching, Germany. E--mail: pilar@MPA--Garching.MPG.DE}

\slugcomment{{\it Running title:} White dwarf halos}

\begin{abstract}
The suggestion that roughly half the mass of the galactic halo might be in
the form of white dwarfs, together with the limits on the infrared background
light and on the initial metallicity of the galactic disk, would set strong
constraints on the initial mass function (IMF) of the halo. Particular IMFs
have been proposed to cut off both the numbers of low--mass stars
contributing to the infrared background and of high--mass stars which
contribute to the growth of metallicity when they promptly explode as
gravitational--collapse (Type II and Type Ib/c) supernovae. Here we examine
the further contribution to metallicity from the Type Ia (thermonuclear)
supernovae which would later be produced from the halo white dwarf
population. We find that, for most of the evolutionary scenarios for the Type
Ia supernova progenitor systems proposed so far, the constraints on the white
dwarf mass fraction in the halo from the predicted production of iron would
be extremely severe. When the predicted iron excess is not so large, then the
exceedingly high Type Ia supernova rate predicted for the present time would
also exclude a major contribution of white dwarfs to the halo mass. The white
dwarf contribution, in all cases, should be below $5-10\%$. Besides, for the
IMFs considered, the duration of the halo burst should be shorter than 1 Gyr
in order to avoid too large a spread in the iron abundances of Population II
halo dwarfs, and the predicted halo $[O/Fe]$ ratio would be at odds with
observations.
\end{abstract}


\section{Introduction}
Microlensing experiments (Bennet et al. 1996; Alcock et al. 1996) might
indicate that roughly half of the mass in the halo of our Galaxy could be
made of white dwarfs (WDs). The existence of such large numbers of WDs, the
remnants of an earlier generation of halo stars, poses different problems.
One of them is that as shown, for instance, by Adams \& Laughlin (1996), the
initial mass function (IMF) of the parent population of those WDs should have
been very different from the IMF inferred for the galactic disk (see also
Chabrier, Segretain, \& M\'era 1997; Fields, Mathews, \& Schramm 1997).
Otherwise, the low--mass tail of the IMF would give red dwarfs much in excess
of their maximum allowed mass fraction in the halo (Graff \& Freese 1996)
while the high--mass end, by its metal production, would raise the initial
metallicity Z of the galactic disk much above any reasonable level (unless
the supernova products were ejected into the intergalactic medium). Far too
luminous galactic halos (which should be seen at high redshifts) would also
result from the large numbers of massive stars. From that double constraint,
Adams \& Laughlin (1996) deduce that the IMF of the WD progenitors should be
confined within the mass range $1 M_{\odot}\lapprox M\lapprox 8 M_{\odot}$,
and be sharply peaked about a characteristic mass $M_{C}\simeq 2.3
M_{\odot}$. Even with such a IMF, due to the fact that only a fraction of the
initial mass of the progenitor star stays trapped in the remnant WD, those
authors (see also Isern et al. 1997) conclude that most likely the WD
contribution to the halo mass should be 25\% or less, 50\% being an extreme
upper limit. More recently, Gibson \& Mould (1997) have examined the
production of C, N, and O by the intermediate--mass star progenitors of the
WDs. They find that the expected [C,N/O] ratios would be hard to reconcile
with those measured in Population II halo dwarfs. Difficulties with models of
WD--dominated halos are also pointed out by Venkatesan, Olinto, \& Truran
(1997). Earlier, Charlot \& Silk (1995) had set upper limits to the WD
fraction in galactic halos from the absence of the luminosity signature of
the WD progenitors in deep galaxy surveys. It should be stressed that mass
determinations from gravitational microlensing are still uncertain (Mao \&
Paczy\'nski 1996; Venkatesan, Olinto, \& Truran 1997). Thus, possible
incompatibilities with observational constraints must be taken into account
before concluding that star formation in the halo should have been very
different from that inferred for the disk.

In this {\it Letter} we look further into the problem of the metal enrichment
of the galactic disk and halo by the putative parent population of the WDs.
Massive stars ($M\gapprox 8-10 M_{\odot}$) eject metals mostly at the end of
their lives, when they explode as supernovae due to the gravitational
collapse of their dense, fuel--exhausted cores. Phenomenologically, those are
Type II (hydrogen--rich) and Type Ib/Ic (hydrogen--devoid) supernovae. Their
progenitors are basically eliminated by adopting the IMF proposed by the
aforementioned authors. However, the very population of halo WDs should give
rise to another type of supernovae: thermonuclear supernovae
(phenomenologically Type Ia, lacking hydrogen in their spectra). Type Ia
supernovae (SNe Ia) are the explosions of some WDs (among those made of C+O)
which ignite when they are compressed by mass accretion from a close binary
companion. Those explosions yield an average of $\sim 0.6\ M_{\odot}$ of
iron, and it is estimated that they produce about 2/3 of the iron in the
galactic disk (Bravo et al. 1993; Woosley \& Weaver 1994), the other 1/3
coming from gravitational--collapse supernovae. Therefore, unless the binary
frequency in the halo were very low and/or the distribution of initial binary
parameters (mass ratios of the two components, binary periods) in the
progenitor population of the halo WDs would strongly suppress the formation
of close binary systems containing C+O WDs, one should expect a large
contribution to the iron contents of the disk from a massive WD halo.

In the following we derive the time evolution of the iron mass produced by
SNe Ia after an initial burst of star formation able to generate the
presumptive WD halo population without violating neither the red dwarf nor
the high--mass stars constraints. Whereas the gravitational--collapse
supernova rate can just be set equal to the massive star formation rate, the
thermonuclear supernova (SNe Ia) rate depends on the evolutionary path
assumed to produce the supernovae from a fraction of the close binary systems
containing C+O WDs, together with the adopted distributions of initial binary
parameters. We will thus discuss the dependence from those hypotheses of the
iron production constraint on the mass fraction of WDs in the halo. We will
see that in most cases the iron enrichment from SNe Ia would be incompatible
with any substantial contribution of WDs to the halo dark matter. In the
remaining cases, the exceedingly high SNe Ia rate predicted for the present
time does also indicate that the mass fraction of the galactic halo in the
form of WDs should be much smaller than that suggested from microlensing
experiments. A further restriction to the hypothesis of particular IMFs
giving rise to a large halo WD population comes from the comparison of the
predicted spread in the iron abundances and the inferred $[O/Fe]$ ratios for
Population II halo dwarfs with observational data.

\section{Modeling, Results, and Discussion}
We will assume that the parent population of the halo WDs forms in a burst
lasting $\sim$ 1 Gyr, with a IMF of the form:

$$ln f(ln\ M) = A - {1\over 2\langle\sigma\rangle^{2}}\left[ln\left({M\over
M_{c}}\right)\right]^{2}\eqno(1)$$

\noindent
where $A$, $M_{c}$, and $\langle\sigma\rangle$ are constants (Adams \&
Laughlin 1996). $A$ sets the total mass in the burst, and for $M_{c}$, the
mass scale of the distribution, and $\langle\sigma\rangle$, its dimensionless
width, the values $M_{C} = 2.3 M_{\odot}$ and $\langle\sigma\rangle = 2.3$
are adopted. The IMFs proposed by Chabrier, Segretain, \& M\'era (1997) and
by Fields, Mathews, \& Schramm (1997) are similar and, to our present
purpose, give equivalent results.

A fraction of the halo stars will be in binary systems whose initial
parameters (primary mass, secondary/binary mass ratio, and separation between
the two components) imply that they should eventually end up as a C+O WD plus
a close companion. Mass transfer from the companion to the WD can then lead
to explosive C ignition and a SNe Ia. Different scenarios have been proposed,
depending on the nature of the companion (another C+O WD, a He star, a
subgiant or red--giant star). We have considered all of them in order to
calculate the SNe Ia rates and corresponding iron production following an
outburst of star formation. Namely, the scenarios are: a) merging of a couple
of C+O WDs (double--degenerate merging: DD); b) explosive ignition of He
(followed by central ignition of C) at the surface of a C+O WD as a result of
accretion from a He star companion (helium cataclysmic variable: HeCV); c)
central explosive ignition of C as a result of mass growth by accretion from
a red--(sub)giant companion (cataclysmic--like system: CLS); d) same as the
previous case, but allowing higher mass--loss rates from the companion
(``wind solution'': CLS(W)); e) explosive ignition of He (produced from
burning of H) at the surface of a C+O WD accreting mass from the wind of a
red--giant or supergiant companion (symbiotic system: SS). In scenarios a),
c), and d) the exploding WD has reached the Chandrasekhar mass while in
scenarios b) and e) the explosion takes place when a thick enough He layer
has accumulated, the C+O WD mass being still below the Chandrasekhar mass.
Most scenarios were already proposed in a seminal paper by Iben \& Tutukov
(1984) (see Iben 1997, for a recent review). The ``wind solution'' for the
CLS scenario has been proposed by Hachisu, Kato, \& Nomoto (1996). The
characterization of the different SNe Ia scenarios considered here is as in
Ruiz--Lapuente, Burkert, \& Canal (1995) and Canal, Ruiz--Lapuente, \&
Burkert (1996) (see also Ruiz--Lapuente, Canal, \& Burkert 1997). The results
reported here correspond to the distributions of initial binary parameters
adopted in those papers. We have also explored other suggested distributions,
but the outcome was not significantly altered.

In Figure 1 we show the growth of the iron mass, $M_{Fe}$, following the halo
star formation outburst, for scenarios a)--e). We will first assume that all
the iron produced by the SNe Ia in the halo directly mixes with the gas in
the disk and we will later discuss the possibilities of relaxing this
hypothesis. The halo mass adopted is $M_{Halo} = 10^{12}\ M_{\odot}$ (its
dynamical mass: see Peebles 1995). We take the disk mass $M_{Disk} =
0.1\times M_{Halo}$. The relevant quantity, four our purpose, is the ratio
$M_{Fe}/M_{Disk}$, that is the $Z_{Fe}$ in the disk due to the halo
contribution. As we see in Figure 1, in all scenarios $M_{Fe}$ grows with
time until reaching a maximum value at an epoch which corresponds to the
explosion of the most slowly evolving SNe Ia progenitors formed in the halo
outburst. The plateau value and the time at which it is reached depend on the
scenario. As a conservative upper limit to the initial $Z_{Fe}$ in the
galactic disk we take half its solar value, $Z_{Fe \odot} = 1.7\times
10^{-3}$ (Cameron 1982). Thus, $Z_{Fe}(0)\lapprox 8.5\times 10^{-4}$. For our
disk mass, that corresponds to $M_{Fe} = 8.5\times 10^{7}\ M_{\odot}$. Such a
value is shown by the dashed horizontal line in the different panels of
Figure 1.

We see that, unless most of the iron produced by SNe Ia from the halo WD
population would not fall into the disk, in three of the scenarios considered
(HeCV, CLS, and CLS(W)) the upper limit would already be reached at the end
of the outburst (and even before that in the HeCV scenario), the iron mass
rapidly growing up to much larger values afterwards. In the SS scenario, the
limit would be reached at $t\simeq 0.5\ Gyr$ after the end of the burst.
Later on, much larger iron masses are produced (it must be noted that the SNe
Ia rates have been very conservatively calculated here: the SS efficiency in
producing SNe Ia has been set to its lowest estimate). Only in the DD
scenario the upper limit would not be reached until $t\simeq 1.5\ Gyr$ after
the end of the burst, to slowly grow up to $\simeq 1.7$ times that, $10.5\
Gyr$ later. The reason for the comparatively low SNe Ia rates in this
scenario is that most of the progenitors of the DD mergers are within the
initial mass range $6 M_{\odot}\lapprox M\lapprox 8 M_{\odot}$, which is not
favored by a IMF of the form (1). Note that, for longer delays between halo
and disk formation, the initial iron abundances in the disk would be higher.

In fact, the halo SNe Ia contribution to the iron enrichment of the disk
would still be dominant at the time of birth of the Sun. Therefore, even
neglecting the contribution from the SNe Ia and gravitational--collapse SN in
the disk (and also the decrease in its gas contents due to previous star
formation), the WD mass fraction in the halo should be $\lapprox 5-10\%$ (for
the HeCV, CLS, and CLS(W) scenarios). From the iron argument alone, up to
$\simeq 20\%$ would still be compatible with the SS prediction, and that for
the DD scenario might even fit the microlensing estimate. However, as we will
see next, consideration of the predicted SNe Ia rates for the present time do
set more restrictive bounds.

\noindent
{\it SNe Ia rates.} It is interesting to note that, in the DD and SS
scenarios, SNe Ia should still be exploding in the halo at $t = 13\ Gyr$
after the start of the initial outburst. In the DD scenario, the rate would
be $\nu_{SNe Ia}\simeq 7\times 10^{-3}\ yr^{-1}$, whereas in the SS scenario
it would be $\nu_{SNe Ia}\simeq 2\times 10^{-2}\ yr^{-1}$. Even at $t = 18\
Gyr$ (a halo age suggested by Chabrier, Segretain, \& M\'era 1997), $\nu_{SNe
Ia}\simeq 4\times 10^{-3}\ yr^{-1}$ in the DD scenario, and $\nu_{SNe
Ia}\simeq 1\times 10^{-2}\ yr^{-1}$ in the SS one. All those values are far
above observational upper limits. The only supernova observed in our Galaxy
in the last 1,000 yr which clearly was a SNe Ia has been SN 1006 (van den
Bergh \& Tammann 1991), and its progenitor did not belong to the halo
population. The argument that the historical supernova sample is not complete
beyond a distance $\sim 3\ kpc$ applies to the thick disk population at most,
but not to typical halo objects. Even at a distance $d\sim 20\ kpc$ (that is
far out in the halo from our location within the Galaxy) the apparent blue
magnitude of a SNe Ia at maximum would still be $m_{B}\simeq -2.5$ (about 1
mag brighter than Sirius). From this argument, a reasonable upper limit to
the halo SNe Ia rate should be at least one order of magnitude below that
predicted by the DD scenario, which would again put the contribution of the
WD population to the halo mass below $5-10\%$. Consideration of the rate of
merging of WD pairs in the halo population strengthens this conclusion, since
it should occur at rates $\nu_{merg}\simeq 0.4\ yr^{-1}$ (13 Gyr) or
$\nu_{merg}\simeq 0.3\ yr^{-1}$ (18 Gyr). Most of them would not produce any
SNe Ia but a hot WD, which would remain very luminous for $\sim 10^{8}\ yr$.
Their total population should thus be $\sim 3-4\times 10^{7}$ nowadays and it
would hardly have escaped detection.

\noindent
{\it The halo gas component.} In the preceding we have assumed that most of
the material ejected by the SNe Ia mixes with the gas in the galactic disk.
If the halo were left completely gas--free after the initial burst of star
formation, one should expect that roughly half the SNe Ia ejecta would escape
the Galaxy whereas the other half would hit the disk and mix with the gas
there (typical velocities of the ejecta are larger than the escape velocity
from the galactic halo and there would not be halo gas to slow them down).
However, as pointed out by Adams \& Laughlin (1996), Isern et al. (1997), and
Gibson \& Mould (1997), the intermediate--mass star progenitors of the WDs
should return typically $\simeq 50-75\%$ of their mass to the interstellar
medium through stellar winds and planetary nebula ejection. The time scale of
such mass ejection is shorter than that of growth of the iron mass from SNe
Ia (especially in the case of the DD and SS scenarios). Therefore, much of
this gas should mix with the iron--rich SNe Ia ejecta. In that case, the iron
would be much more diluted than if it were only mixed with the gas in the
disk. There is, however, the problem of where this halo gas might be hidden
(it cannot fall into the disk, since otherwise the latter would be much more
massive than it actually is). One possibility is that it would be
concentrated into cold molecular clouds (De Paolis et al. 1997). In that
case, since formation of most clouds would precede ejection by the SNe Ia of
a major fraction of the iron mass and the filling factor of the clouds should
be small, the case would resemble that of a gas--free halo. Another
possibility is that heating of the halo gas by the SNe Ia explosions might be
enough to generate a strong galactic wind and eject most of the gas. A first
estimate of the energy budget shows that this possibility is only marginal.
Summarizing, in any case the problem of the halo gas component, rather than
rising the upper limit to the WD mass in the galactic halo, strengthens the
conclusion that it should be smaller than the total mass in the galactic disk
and make $\lapprox 5-10\%$ of the halo mass only.

\noindent
{\it Halo dwarf metallicities and $[O/Fe]$ ratios.} As it can already be seen
in Figure 1, even within a burst of star formation lasting $\sim 1\ Gyr$,
there should be an appreciable variation in the iron contents of the gas from
the beginning to the end of the burst, due to the SNe Ia exploding during
this time interval. That should translate into a range of abundances $[Fe/H]$
within the halo Population II dwarfs. The predicted range will depend on the
duration of the burst, on the time variation of the star formation rate
within the burst, and on the SNe Ia scenario. In Figure 2 we compare the
growth with time of the iron abundance for the stars formed in the halo burst
(lasting 1 Gyr), among the scenarios a)--e) considered above, and assuming a
constant star formation rate. We see that, for any scenario, a significant
spread in the iron abundances among halo dwarfs should be expected. A
fraction of dwarfs would show near--solar iron abundances (and even higher,
for the HeCV scenario). The measured spread is $-3.5\lapprox [Fe/H]\lapprox
-1$ (Schuster \& Nissen 1989). Agreement could be obtained in all cases by
shortening the duration of the burst, but that would be in conflict with
evidence of a scatter $\sim 2-3\ Gyr$ in the ages of those stars ({\it
ibid.}). A more difficult problem is that SNe Ia explosions alone would
always give $[O/Fe]\simeq -1.4$ ($-1$ being an extreme upper limit) while the
observed $[O/Fe]$ in halo dwarfs is $\gapprox +0.5$ (and thus bears the
signature of massive star nucleosynthesis) (Barbuy 1988; Abia \& Rebolo 1989;
Spite \& Spite 1991; Spiesman \& Wallerstein 1991). Therefore, not only the
duration of the halo burst should be extremely short to avoid contamination
by the SNe Ia products, but in addition some even earlier generation of
massive stars should have produced the $[O/Fe]$ actually measured. That
appears unlikely.

\section{Conclusions}

If a large fraction of the galactic halo mass were made of WDs, one would
expect a large iron production from the SNe Ia arising from a fraction of the
WDs belonging to close binary systems. Since both the total iron yield and
its time evolution depend on the evolution assumed for the SNe Ia
progenitors, we have considered all the main SNe Ia scenarios proposed so
far: double degenerate merging, He--star cataclysmic systems, two different
versions of the cataclysmic--like scenario, and symbiotic systems. The
results for each evolutionary path are fairly insensitive to the choices of
the initial binary parameters (distribution of mass ratios of the secondary
to the primary, distribution of initial separations). The IMF is chosen to
minimize the numbers of both red dwarfs and high--mass stars.

Assuming that most of the iron produced by the SNe Ia in the halo mixes with
the gas in the disk, the constraint that $Z_{Fe}$ at the time of birth of the
Sun should not exceed $Z_{Fe \odot}$ sets upper limits $\simeq 5-10\%$ to the
WD mass fraction in the halo for the HeCV, CLS, and CLS(W) scenarios.
Comparison of the predicted SNe Ia rates for the present time with
observational upper limits set similar bounds for the SS and DD scenarios.
Besides, a halo burst with IMFs such as those tested here would produce too
large a spread in the iron abundances of Population II halo dwarfs unless it
lasted less than 1 Gyr, which would conflict with evidence of a wider range
of ages among those stars. Worse still, the O/Fe ratio should be far below
solar while the measured ratio is much larger than solar, as one would expect
from massive star nucleosynthesis, and that hardly fits with the proposed
halo IMFs.

Consideration of the role that the SNe Ia originated in the halo WD
population would play in the evolution of the Galaxy, thus points in the same
direction as that of the fate of the gas ejected by the WD progenitors (Adams
\& Laughlin 1997; Isern et al. 1997) and that of the [C,N/O] ratios which
would result (Gibson \& Mould 1997): the WD mass fraction in the galactic
halo should be much lower than that suggested from microlensing experiments.
Our derived upper bounds are even lower than those set from number counts in
deep galaxy surveys.

As a more general conclusion, we can say that the proposal of {\it ad hoc}
IMFs to explain a presumptive huge halo WD population poses problems which
could only be solved by assuming that the whole process of star formation
(single as well as binary stars) in the galactic halo has been completely
different from all we know from local observations. Since the existence of a
massive WD halo is by no means firmly established, it is premature to make so
many {\it ad hoc} assumptions.

\clearpage

\begin{figure}[hbtp]
\centerline{\epsfysize15cm\epsfbox{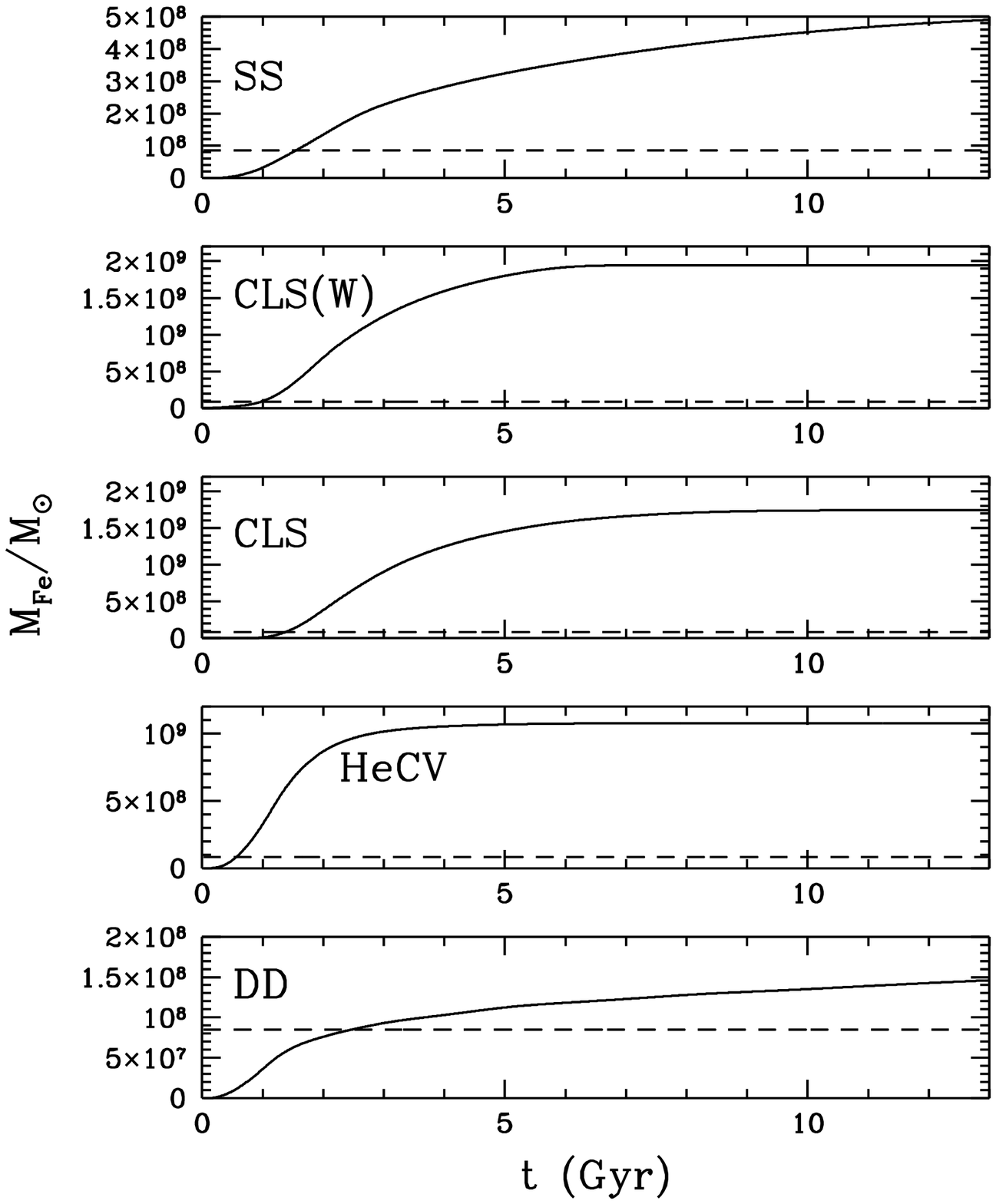}}
\nopagebreak[4]
\figcaption{Growth of the iron mass produced by SNe Ia following the start of
a burst of star formation in the halo, lasting for $1\ Gyr$ and involving
$10^{12}\ M_{\odot}$, for the five different SNe Ia scenarios considered (see
text). The horizontal dashed line corresponds to $M_{Fe} = 8.5\times 10^{7}\
M_{\odot}$, the amount which mixed with $M_{Disk} = 10^{11}\ M_{\odot}$ of
unenriched gas would give $Z_{Fe} = {1\over 2} Z_{Fe \odot}$. Note the
different scales from one panel to another. \label{fig1}}
\end{figure}

\begin{figure}[hbtp]
\centerline{\epsfysize15cm\epsfbox{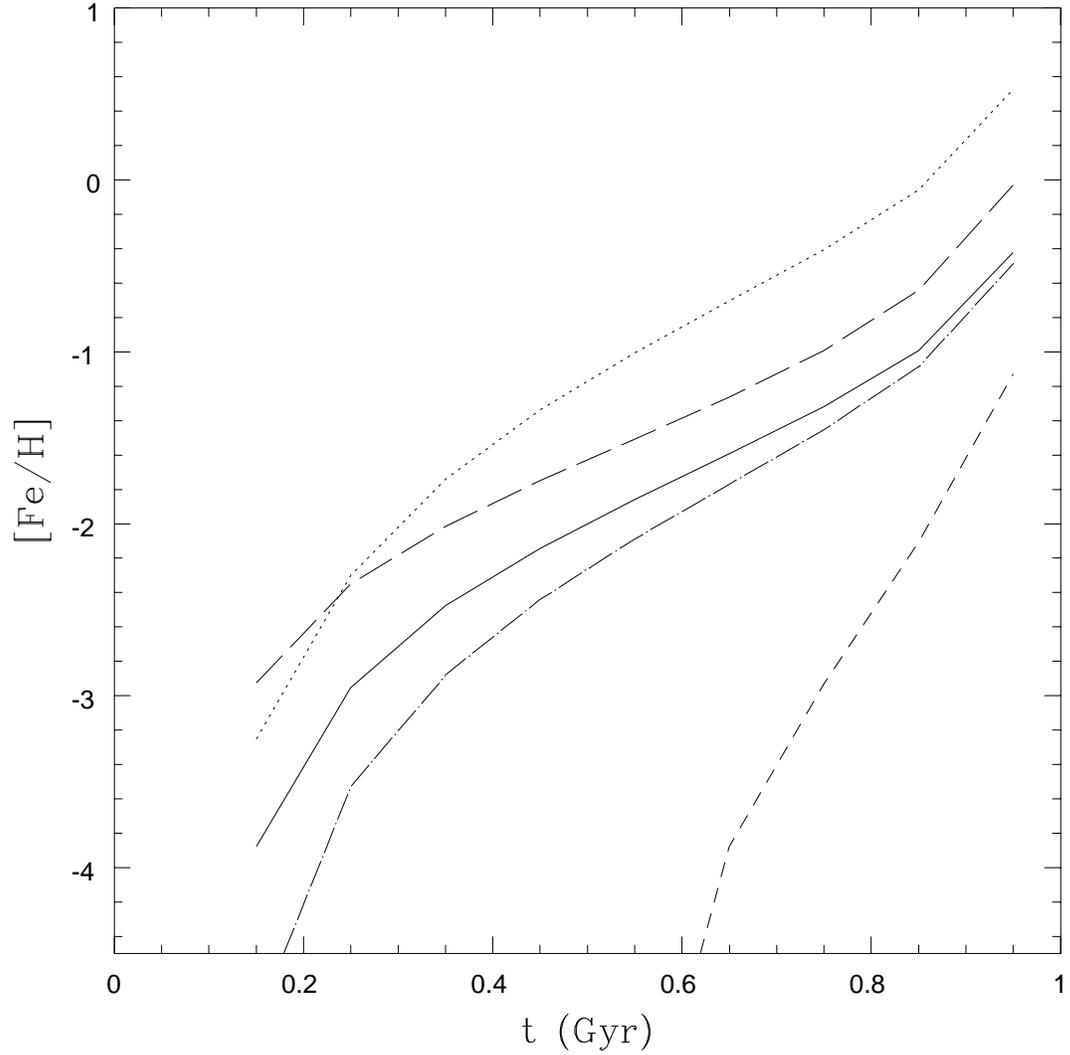}}
\nopagebreak[4]
\figcaption{Growth with time, during the burst, of the iron abundance in
Population II halo dwarfs, for scenarios a)--e), assuming a constant star
formation rate. Continuous line: DD; dotted line: HeCV; short--dashed line:
CLS; long--dashed line: CLS(W); dot--dashed line: SS (see text and Figure 1).
\label{fig2}}
\end{figure}

\end{document}